\documentclass[a4paper,12pt]{article}

\usepackage{amsmath}
\usepackage{graphicx}
\usepackage{cite}

\title{Effect of Foregrounds on the CMBR Multipole Alignment}
\author{Pavan K. Aluri$^1$, Pramoda K. Samal$^2$,\\ Pankaj Jain$^1$ and John. P. Ralston$^3$}
\date{}

\begin{document}
\maketitle
\begin{center}
{$^{1}$Department of Physics, Indian Institute of Technology, Kanpur 208016, India \\ }
{$^2$Department of Physics, Utkal University, Bhubaneswar, 751004, India\\}
{$^3$}  Department of Physics \& Astronomy, University of Kansas,\\ Lawrence, KS - 66045, USA\\ 
\end{center}

\begin{abstract}
We analyze the effect of foregrounds on the observed alignment of CMBR
quadrupole and octopole. The alignment between these multipoles is studied
by using a symmetry based approach which assigns a principal eigenvector
(PEV) or an axis with each multipole. 
We determine the significance of alignment between 
these multipoles by using the Internal Linear Combination (ILC) 5 and 7 year maps and also the maps obtained
by using the Internal Power Spectrum Estimation (IPSE) procedure. 
The effect of foreground cleaning is studied 
in detail within the framework of the IPSE method both analytically
and numerically. By using simulated 
CMBR data, we study how the PEVs of the pure simulated CMB map differ from 
those of the final cleaned map. We find that, in general, the shift in the 
PEVs is relatively small and in random directions. Due to the random
nature of the shift we conclude that it can only lead to misalignment
rather than alignment of multipoles. We also directly estimate the 
significance of alignment by using simulated cleaned maps. We find that
the results in this case are identical to those obtained by simple
analytic estimate or by using simulated pure CMB maps.     
\end{abstract}


\section{Introduction}

The standard cosmological model rests on the \emph{Cosmological Principle}
which states that the universe is homogeneous and isotropic. However, there
are many indications from diverse data sets that this principle may
not be applicable. In particular, polarizations
of radio waves coming from distant radio galaxies (Birch 1982; Jain and
Ralston 1999),
the optical polarizations from quasars (Hutsem\'{e}kers 1998; 
Hutsem\'{e}kers and Lamy 2001, Jain et al. 2004)
and the multipoles $l=1,2,3$ of Cosmic Microwave Background Radiation (CMBR) 
all indicate a universal axis pointing in the direction of the Virgo
cluster (de Oliveira-Costa  et al. 2004; Ralston and Jain 2004;
Schwarz et al 2004). This phenomenon has been
called the \emph{Virgo Alignment} (Ralston and Jain 2004). 
There have also been other claims of large scale
anisotropy in the CMBR data. These include the dipole power modulation
(Eriksen et al. 2004; Eriksen  et al. 2007; Hoftuft  et al. 2009), 
as well as, the detection of an
anomalously cold spot (Cruz et al. 2005).  
These anomalies have been studied in great detail 
(Bielewicz et al. 2004; Hansen et al. 2004;
Katz and Weeks 2004; Bielewicz  et al. 2005;
Prunet et al 2005; Bernui  et al. 2006; Copi et al. 2006;
Copi et al. 2007; de Oliveira-Costa and Tegmark 2006; Wiaux et al. 
2006; Freeman et al. 2006;
Helling et al. 2006; Bernui et al. 2007; Land and Magueijo 2007;
Magueijo and Sorkin 2007). 
Meanwhile some studies do not find any violation of statistical isotropy 
in CMB (Hajian et al. 2005; Hajian and Souradeep 2006).
Furthermore, the cluster
peculiar velocities (Kashlinsky et al. 2008, 2009), as well as the large 
scale galaxy distribution (Itoh et al. 2009), also
indicate deviations from isotropy. The preferred
axis for the cluster peculiar velocities with a high redshift cut is again
found to be close to the direction of Virgo cluster. In the galaxy
distribution surveys, the preferred axis appears to depend strongly on
the cut made on the data.

There have been a variety of proposals in the literature explaining
the possible origin of this \emph{alignment anomaly}, such as,
anisotropic space-times (Berera  et al. 2004; Kahniashvili et al. 2008),
foreground contaminations (Slosar and Seljak 2004; Abramo et al. 2006; Rakic et al. 2006), 
noise bias or systematics (Naselsky et al. 2008), 
vector like dark energy (Armendariz-Picon 2004), spontaneous breaking
of isotropy (Gordon et al.  2005; Land and Magueijo 2006, Erickcek et al. 
2009), 
inhomogeneous universe (Moffat 2005; Land and Magueijo 2006), inhomogeneous 
inflation (Carroll et al. 2010), violation of rotational invariance during inflation (Ackerman et al. 2009), 
anisotropic perturbations due to dark energy 
(Battye and Moss 2006), anisotropic inflation
(Buniy et al. 2006), local voids (Inoue and Silk 2006) and
anisotropic dark energy (Koivisto and Mota 2008; Rodrigues  2008).


Another interesting effect associated with quadrupole is its low power 
in comparison to the best fit LCDM model (Bennett et al. 2003). One may 
speculate that 
the observed quadrupole-octopole alignment is related to the observed
low quadrupole power. It has been shown that in random realizations of
pure CMB maps there is no correlation between
the low quadrupole power and quadrupole-octopole alignment 
(Raki\'{c} and Schwarz 2007; Sarkar et al. 2010).

In the present paper, we analyze the alignment of CMBR quadrupole $(l=2)$
and octopole $(l=3)$ moments more closely. 
In (Slosar and Seljak 2004), it
has been suggested that this alignment may be caused by the galactic
foregrounds. These authors argue that most of the power of quadrupole
and octopole lies in the most contaminated region of the sky. Hence, it is
entirely possible that these multipoles may be significantly contaminated
and the alignment may be caused due to poor cleaning. 

In the Internal Power Spectrum Estimation (IPSE) procedure
(Saha et al. 2006, 2008; Samal et al. 2010) the cleaning is performed 
by a totally blind procedure without making any explicit model for 
foregrounds. The foregrounds are removed by adding maps, in harmonic space,
 at different
frequencies with suitable weights, chosen so as to minimize the 
foreground power. The Internal Linear Combination (ILC) map provided by 
the WMAP team (Bennett et al. 2003) also uses a similar cleaning procedure.
Here the weighted maps are added directly in pixel space. Both of these 
procedures are highly effective in removing foregrounds. However 
some residual foregrounds may still contaminate the data. 
Besides this, (Saha et al. 2008; Samal et al. 2010)
pointed out the existence of bias in the low $l$ multipoles 
extracted from foreground cleaned maps. This
bias leads to a significant reduction of extracted power in these multipoles.
In fact, the bias completely explains the observed low power at $l=2$, in comparison
to the best fit theoretical LCDM model. It is clearly important to
determine what the low $l$ bias might imply for the alignment of these
multipoles. It is possible that the power gets eliminated in just the precise
manner in order to cause alignment. 
Here we study whether any of these 
residual effects, present in the foreground cleaned data, may cause the
observed quadrupole-octopole alignment. 

It is important to
note that the observed alignment is only getting better with more data.
A larger data sample implies smaller uncertainties due to detector noise. 
Hence the signal becomes more prominent as the fluctuations due to 
detector noise get suppressed.

The problem of testing violation of isotropy in the CMBR signal has been
addressed by many authors. The CMBR anisotropy spectrum is generally assumed
to be isotropic in a statistical sense. The assumption of statistical isotropy
of the CMBR signal means that the spectral coefficients, $a_{lm}$, are
uncorrelated for different $m$ and $l$, i.e. the ensemble average,
\begin{equation}
<a^*_{lm} a_{l'm'}> = \delta_{ll'} \delta_{mm'} C_l\,,
\end{equation} 
where $C_l$ is the standard power. A multitude of statistics have been
proposed in order to test for deviation from statistical isotropy in the CMBR
data (Hajian et al. 2005; Hajian and Souradeep 2006; Copi et al. 2006, 2007). Here we shall use the method introduced in (Ralston and Jain 2004)
and further developed in (Samal  et al. 2008, 2009). In this
procedure, one assigns three orthogonal unit vectors for each multipole. The
orientation of these vectors as well as the power associated with each vector
contains information about possible violation of statistical isotropy. This
information is encoded in two entropy measures, the power entropy and alignment
entropy, defined in (Samal  et al. 2008). Using this method one can test for
violation of isotropy for each individual multipoles or collectively over a
range of multipoles. We present an outline of our analysis procedure in the
next section followed by results from cosmological data and simulations and
conclusions in the end.

\section{Analysis Methodology}
The primary objective of the present paper is to determine whether the 
alignment of quadrupole and octopole can be caused by foreground contamination.
The alignment of different multipoles can be quantified by associating
a frame with each multipole (Ralston and Jain 2004; Samal  et al. 2008). In order to analyse
the effect of foreground contamination we first perform an analytic 
calculation of the power tensor, introduced in (Ralston and Jain 2004; Samal 
 et al. 2008), 
within the framework of  IPSE 
technique (Saha et al. 2006, 2008; Samal et al. 2010). Here we confine ourselves to
the simple case of one foreground component following rigid frequency
scaling and two frequency bands. This computation is useful to
determine if the low $l$ bias also leads to violation of statistical
isotropy. We then numerically determine the effect of this bias and residual
foregrounds on the 
alignment of quadrupole and octopole. We generate  
many random realizations
of the CMBR maps including foregrounds and detector noise.
We use the Planck Sky Model (PSM)\footnote[3]{http://www.planck.fr/heading79.html} for foregrounds and use all the 
five WMAP frequency bands.
 These simulated raw
maps are cleaned using the IPSE technique. We study the difference in the
extracted principal eigenvectors of cleaned maps and 
simulated pure CMB maps. 
 This allows us to determine if the
residual foregrounds lead to 
any preferential alignment of these extracted vectors. 

In the next subsections, we briefly review
the extraction of frames and the IPSE technique.

\subsection{Covariant frames}

The present study utilizes a symmetry based statistic (Samal et al. 2008) 
to test for statistical isotropy in the CMBR data. The method associates a wave
function $\psi^k_m$, defined as,
\begin{equation}
\psi^k_m(l) = { 1 \over {\sqrt{l(l+1)}} } \langle l,m|J^k|a(l) \rangle
\label{eq:wavefunction}
\end{equation}
with every multipole $l$. Here $|a(l) \rangle $ contains information about the
spectral moments, such that,
\begin{equation}
a_{lm} = \langle l,m|a(l) \rangle
\end{equation}
where $|lm \rangle $ are the eigenstates of the angular momentum operators
$\vec J^2$, $J_z$ in the spin-$l$ representation. The wave function $\psi^k_m$
is useful since it assigns a frame, i.e. three orthonormal vectors, at each $l$.
The frame may be extracted by making a singular value decomposition of the
wave function.

It is also convenient to define the power tensor, 
\begin{eqnarray}
 A_{ij}(l) &=& \underset{m}{\sum} \psi^i_m {\psi^j_m}^* \nonumber\\
&=& {1\over l(l+1)} \sum_{m,m'} a^*_{lm}(J_iJ_j)_{mm'}a_{lm'}\, .
\end{eqnarray}
The three eigenvectors, $e^\alpha_i$, $\alpha = 1,2,3$, of this matrix
define the frame associated with each multipole. The index, $i=1,2,3$, labels
the three components of each eigenvector. The corresponding eigenvalues,
$(\Lambda^\alpha)^2$, contain information about the power associated
with each eigenvector. The sum of the three eigenvalues equals the total 
power, $C_l$. A large dispersion in the eigenvalues indicates
violation of statistical isotropy for a particular mode. 

We define the principal pigenvector (PEV), $\hat n_l$, as the eigenvector
corresponding to the maximum eigenvalue. Here we note that the extracted PEVs 
are headless, meaning, these vectors
specify only an axis and not a direction. The P-value or the
significance of alignment of the
quadrupole and octopole PEV may be estimated analytically by the formula
$P = 1-\hat n_2\cdot \hat n_3= 1-\cos(\delta\Theta)$. Here $\delta\Theta$ is
the angle between the PEVs $\hat n_2$ and $\hat n_3$, which correspond
to $l=2$ and $l=3$ respectively. 
Alternatively, one can estimate the significance directly by numerical
simulations, as described below.

\subsection{IPSE method}
The IPSE procedure removes the foreground contamination by linearly
combining maps at
different frequencies with suitable weights in harmonic space. 
The cleaning is accomplished independently for each $l$. 
Let  $\hat w_l^a$ denote the weights for the map at frequency channel $a$
corresponding to the multipole $l$. In case we have several maps
at a particular frequency, we simply take their average. 
The spherical harmonic components of the cleaned map are given by,
\begin{equation}
a_{lm}^{\rm Clean}=\sum_{a=1}^{n_c} \hat w_l^{a}\frac{a_{lm}^a}{B_l^a} \, .
\label{c_map}
\end{equation}
Here $n_c$ is the total number of frequency channels used for cleaning. 
The factor $B_l^a$ is the
circularized beam transform function at frequency band $a$ (Hill et al.
2008). 
The $a_{lm}^a$ represent the harmonic coefficients of the observed map. 
It can be expressed as
\begin{equation}
a^a_{lm} = (a^s_{lm} + a^{(f)a}_{lm}) B_l^a + a^{(N)a}_{lm}
\end{equation}
where $a^s_{lm}$, $a^{(f)a}_{lm}$ and $a^{(N)a}_{lm}$ represent the
contributions from CMB, foregrounds and detector noise respectively
at frequency channel $a$. 

The weights $\hat w_l^a$ are obtained by minimizing the
total power subject to the constraint
\begin{equation}
 {\hat {\bf W}}_l{\bf e}_0={\bf e}^T_0{\hat{\bf W}}^T_l=1 \, ,
\label{constraint}
\end{equation}
where
 ${\bf e}_0$  is a column vector with unit elements
 \begin{equation}
 {\bf e}_0 =\left(
 \begin{array}{c}
 1  \\
 ..  \\
 ..  \\
 1
 \end{array}
 \right) \, ,
 \end{equation}
and $\hat {\bf W}_l$ is the row vector $(\hat w_l^1,\hat w_l^2,..,\hat w_l^{n_c})$. This gives
\begin{equation}
 \hat{\mathbf{W}}_l = {\mathbf{e}_0^T \hat{\mathbf{C}}_l^{-1} \over \mathbf{e}_0^T \hat{\mathbf{C}}_l^{-1}\mathbf{e}_0}\,\,,\,\,
\end{equation}
where $\hat{\bf C}_l$ is the empirical covariance matrix
(Tegmark and Efstathiou 1996; Tegmark et al. 2003; 
Saha  et al. 2006;  Delabrouille and Cardoso 2009). 
Its $ab$ matrix element may be expressed as,
$\hat C^{ab}_l/(B_l^a B_l^b)$, where $\hat C_l^{ab}$ is the cross power 
spectrum between the $a^{th}$ and $b^{th}$ channel,
\begin{equation}
\hat C^{ab}_l = \sum_{m=-l}^{m=l} {a_{lm}^a a_{lm}^{b*}\over 2l+1}\,.
\end{equation}
The final
cleaned map may be used to extract the power spectrum as well as the
PEVs needed for the present study. The power spectrum is
given by,
\begin{equation}
\hat{C}_l^{Clean} = {1 \over \mathbf{e}_0^T \hat{\mathbf{C}}_l^{-1}\mathbf{e}_0}\,.
\end{equation}
This would be reliable
at low $l$, where the detector noise is negligible. For our present purpose,
this is sufficient. However at high $l>200$, where the detector noise is not
negligible, a more elaborate procedure is necessary, as discussed in
(Saha et al. 2008). 

The procedure contains some bias which can be estimated from simulations. An
important bias arises due to inefficient cleaning in the galactic plane. A
similar bias is also present in the procedure used by WMAP science team in
generating the ILC map. 
As a final step in generating this map, a `bias' correction based on
Monte Carlo simulations is applied by the WMAP science
team (Hinshaw et al. 2007). We employ a similar procedure to correct for any residual
foreground bias by subtracting a `bias map' generated from the simulated maps,
in pixel space.

Another interesting bias arises due to the cross correlation term between the
foregrounds and the CMB signal in the calculation of CMB power. 
This bias is negative and affects dominantly
the very low multipoles. This can also be estimated by simulations, but may
also be represented analytically by using some simplifying assumptions 
(Saha  et al. 2008).

\subsection{Simulations}

In order to study the effect of foregrounds on the detected anisotropy in CMBR,
we generate an ensemble of 500 CMB maps as random realizations of the best fit 
theoretical power spectrum, available at the NASA's WMAP public domain
website\footnote[1]{http://lambda.gsfc.nasa.gov/}. These simulated ``pure" CMB
maps are generated using the publicly available HEALPix software\footnote[2]{http://healpix.jpl.nasa.gov/}.
Eventually, these pure CMB maps are contaminated with galactic foregrounds 
and gaussian random noise with appropriate dispersion per pixel. 
The foregrounds are generated using
the PSM, 
which contains only thermal dust, synchrotron and free-free emission. 
These are the dominant foregrounds, as 
characterised by WMAP science team also, in all
its data releases so far (Gold  et al. 2010).
The effective number
of observations needed to generate the noise maps are taken from temperature
maps of WMAP 5 year and 7 year data releases, available in FITS format at 
NASA's LAMBDA website\footnotemark[1]. These simulated CMB maps
are then passed through a cleaning pipeline using the IPSE procedure. This
produces a full sky cleaned map corresponding to each random realization which can
be used to study the alignment between quadrupole and octopole. By comparing observations
with the random samples, we can easily estimate the significance level of the
observed alignment. Furthermore, we can study how the PEVs in
the extracted clean CMB maps differ from those of pure CMB maps. This can reveal
any systematic bias which may be present in the PEVs of cleaned maps.
Due to the presence of residual foregrounds and negative bias at low $l$, we expect that the corresponding
eigenvectors may also be biased. Here we shall study this bias and its effect on
alignment of quadrupole and octopole.

\section{Effect of low-\emph{l} bias}

In this section we evaluate analytically how the low-\emph{l} negative bias 
affects the power tensor. For simplicity we assume that there is only
one significant foreground component, which follows rigid frequency 
scaling. Furthermore we assume only two frequency bands in the CMB data
analysis. The detector noise is assumed to be negligible, which is a
reasonable assumption at low-$l$. For simplicity we also set the beam 
transform function to unity. We compute the ensemble average of
the power tensor within the framework of the IPSE method using these
simplifying assumptions.

The wave function $\psi^i_m(l)$ associated with each multipole, Eq. 
\ref{eq:wavefunction}, may be expressed as, 
\begin{eqnarray}
\psi_m^i(l) 
	    &=& {1 \over \sqrt{l(l+1)}} \underset{m'}{\sum} \langle lm|J_i|lm' \rangle a_{lm'}\ .
\end{eqnarray}
The corresponding power tensor may be written as, 
\begin{eqnarray}
 A_{ij}(l) 
	   &=& {1 \over l(l+1)}\underset{m,m',m''}{\sum} \langle lm|J_i|lm' \rangle \langle lm''|J_j|lm \rangle a_{lm'} a_{lm''}^{*}\,.
\end{eqnarray}
To calculate the bias in the power tensor we take its ensemble average, 
to get,
\begin{equation}
{\langle A_{ij}(l) \rangle}_{ens.avg.} = {1 \over l(l+1)}\underset{m,m',m''}{\sum} \langle lm|J_i|lm' \rangle 
					  \langle lm''|J_j|lm \rangle {\langle a_{lm'} a_{lm''}^{*} \rangle}_{ens.avg.}\,,
\end{equation}
where the symbol $< >_{ens.avg.}$ denotes ensemble average.
It follows from the \emph{Cosmological principle} that the fluctuations in CMBR are statistically
isotropic. Hence we expect for a pure CMB map, 
\begin{equation}
{\langle A_{ij}(l) \rangle}_{ens.avg.} = {C_l\over 3} \delta_{ij}\,.
\end{equation}
Here we compute the ensemble average of the power tensor for the cleaned map.
This will show us whether the
bias generated due to foreground cleaning leads to anisotropy in the
power tensor.

The spherical harmonic coefficients of a cleaned map can be written as,
\begin{equation}
 a_{lm}^{Clean} = a_{lm}^{(s)} + a_{lm}^{res} \,,
\end{equation}
where $a_{lm}^{res}$ denotes any residual foreground contamination in the cleaned map,
\begin{equation}
 a_{lm}^{res} = \sum_{b=1}^{n_c} \hat{w}^b_l a_{lm}^{(f)b}\,.
\end{equation}
In order to estimate the bias in power tensor, we need to calculate,
\begin{eqnarray}
 {\langle a_{lm'}^{Clean}{a_{lm''}^{Clean}}^* \rangle}_{ens.avg.} &=&
{\left\langle \left( a_{lm'}^{(s)} + \sum_{b=1}^{n_c} \hat{w}^b_l a_{lm'}^{(f)b} \right)
{\left( a_{lm''}^{(s)} + \sum_{b'=1}^{n_c} \hat{w}^{b'}_l a_{lm''}^{(f)b'} \right)}^* \right\rangle} \nonumber \\
&=& C_l^{(s)}\delta_{m' m''} + {\langle a_{lm'}^{(s)} \hat{w}^{b'}_l {a_{lm''}^{(f)b'}}^*\rangle} +
    {\langle {a_{lm''}^{(s)}}^* \hat{w}^{b}_l a_{lm'}^{(f)b} \rangle} + \nonumber \\
 && +\,\, {\langle \hat{w}^{b}_l a_{lm'}^{(f)b} {a_{lm''}^{(f)b'}}^* \hat{w}^{b'}_l \rangle}\,.
\label{alm_ens_avg}
\end{eqnarray}
The first term follows from the assumption that CMB signal is statistically isotropic.
The frequency band indices
$b$ and $b'$ in the second, third and fourth terms after second equality are summed over
as in Einstein's convention. 

The elements of the empirical covariance matrix are given by,
\begin{eqnarray}
 \hat{C}_l^{ab} &=& {1 \over 2l+1}\sum_{m=-l}^{+l} a_{lm}^a a_{lm}^{b*} \nonumber \\
		&=& {1 \over 2l+1}\sum_{m=-l}^{+l}\left(a_{lm}^{(s)}+a_{lm}^{(f)a}\right){\left(a_{lm}^{(s)}+a_{lm}^{(f)b}\right)}^* \nonumber \\
		&=& \hat{C}_l^{(s)} + \hat{C}_l^{(s)(f)a} + \hat{C}_l^{(s)(f)b} + C_l^{(f)a(f)b}\,.
\end{eqnarray}
We get the last line after carrying out the multiplications in the second line and
$\hat{C}_l^{(s)(f)a}$ and $\hat{C}_l^{(s)(f)b}$ denote cross correlations
between the CMB and foregrounds at frequency bands $a$ and $b$, respectively. Note that,
these quantities vanish on an ensemble average, as CMB and foregrounds are uncorrelated
with one another i.e., $\langle C_l^{(s)(f)a} \rangle\,=\,0\,=\, \langle C_l^{(s)(f)b} \rangle$.
However, for a particular realization these quantities need not vanish.

As mentioned above we make the estimate of bias in power tensor assuming
 a single foreground component and 2
frequency channels. Thus, the empirical covariance matrix reads as,
\begin{eqnarray}
 \hat{\mathbf{C}}_l = \left(
\begin{smallmatrix}
 \hat{C}^{(s)}_l + 2\hat{C}^{(s)(f)1}_l + C^{(f)1}_l	&	\hat{C}^{(s)}_l + \hat{C}^{(s)(f)1}_l + \hat C^{(s)(f)2}_l + C^{(f)1(f)2}_l \\
 \hat{C}^{(s)}_l + \hat{C}^{(s)(f)1}_l + \hat C^{(s)(f)2}_l + C^{(f)1(f)2}_l	&	\hat{C}^{(s)}_l + 2\hat{C}^{(s)(f)2}_l + C^{(f)2}_l
\end{smallmatrix}
\right)\,.
\end{eqnarray}
 Hence,
\begin{eqnarray}
 \hat{\mathbf{C}}_l^{-1} = {1 \over \hat{\Delta}}\left(
\begin{smallmatrix}
 \hat{C}^{(s)}_l + 2\hat{C}^{(s)(f)2}_l + C^{(f)2}_l	&	-\left( \hat{C}^{(s)}_l + \hat{C}^{(s)(f)1}_l + \hat C^{(s)(f)2}_l + C^{(f)1(f)2}_l\right) \\
 -\left(\hat{C}^{(s)}_l + \hat{C}^{(s)(f)1}_l + \hat C^{(s)(f)2}_l + C^{(f)1(f)2}_l\right)	&	\hat{C}^{(s)}_l + 2\hat{C}^{(s)(f)1}_l + C^{(f)1}_l
\end{smallmatrix}
\right),
\end{eqnarray}
where $\hat{\Delta}$ denotes the determinant of $\hat{\mathbf{C}}_l$. Using the above result
we compute $\mathbf{e}_0^T \hat{\mathbf{C}}_l^{-1}\mathbf{e}_0$ to get the cleaned map power
spectrum as
\begin{eqnarray}
\hat{C}_l^{Clean} &=& {\hat{\Delta} \over {C^{(f)1}_l + C^{(f)2}_l - 2C^{(f)1(f)2}_l}} \nonumber \\
  &=& \hat{C}_l^{(s)} + {1 \over \delta}\left[ 2\hat{C}_l^{(s)(f)1} C_l^{(f)2} + 2C_l^{(s)(f)2}\hat{C}_l^{(f)1}  + C_l^{(f)1}C_l^{(f)2} - \right. \nonumber \\
  && \,-\,(C_l^{(f)1(f)2})^2 - 2C_l^{(f)1(f)2}\left(\hat{C}_l^{(s)(f)1} + \hat{C}_l^{(s)(f)2}\right) \nonumber \\
  && \left.\,-\,(\hat{C}_l^{(s)(f)1})^2 - (\hat{C}_l^{(s)(f)2})^2 + 2\hat{C}_l^{(s)(f)1}\hat{C}_l^{(s)(f)2} \right]
\label{Cl_clean}
\end{eqnarray}
and the corresponding weights,
\begin{eqnarray}
 \hat{\mathbf{W}}_l &=& {\mathbf{e}_0^T \hat{\mathbf{C}}_l^{-1} \over \mathbf{e}_0^T \hat{\mathbf{C}}_l^{-1}\mathbf{e}_0} \nonumber \\
  &=& {1 \over \delta }\left[
 \begin{array}{c}
-\hat{C}_l^{(s)(f)1} + \hat{C}_l^{(s)(f)2} - C_l^{(f)1(f)2} + C_l^{(f)2} \\
 \hat{C}_l^{(s)(f)1} - \hat{C}_l^{(s)(f)2} - C_l^{(f)1(f)2} + C_l^{(f)1}
 \end{array}
  \right]\,\,.
 \label{Weights}
\end{eqnarray}
The term $\delta$ in the denominator of Eq. \ref{Cl_clean} and Eq. \ref{Weights} is
$\delta = C^{(f)1}_l + C^{(f)2}_l - 2C^{(f)1(f)2}_l$. We next calculate 
all the terms in Eq. \ref{alm_ens_avg}
one by one. First, consider the second term in second line
of Eq. \ref{alm_ens_avg}:
\begin{equation}
a_{lm'}^{(s)}\hat{w}_l^{b'} a_{lm''}^{(f)b'*} = a_{lm'} ^{(s)}\hat{w}_l^1 a_{lm''}^{(f)1*} + a_{lm'} ^{(s)}\hat{w}_l^2 a_{lm''}^{(f)2*}
\label{alm_ea_term2-a}
\end{equation}
Using Eq. \ref{Weights} in Eq. \ref{alm_ea_term2-a} and then taking an ensemble average,
we get,
\begin{eqnarray}
 {\langle a_{lm'}^{(s)}\hat{w}_l^{b'} a_{lm''}^{(f)b'*} \rangle} &=& 
{C_l^{(s)} \over (2l+1)\delta}\left[-a_{lm'}^{(f)1} a_{lm''}^{(f)1*} + a_{lm'}^{(f)1} a_{lm''}^{(f)2*} \right. \nonumber \\
&& \left. \,+\, a_{lm'}^{(f)2} a_{lm''}^{(f)1*} - a_{lm'}^{(f)2} a_{lm''}^{(f)2*} \right]\,.
\label{alm_ea_term2-b}
\end{eqnarray}

To proceed further we assume that the foreground component 
follows rigid frequency scaling.
In astrophysical applications, the map of any foreground component at a frequency $\nu$
is generally modelled as,
\begin{equation}
 F(\hat{n}) = A_{\nu_0}(\hat{n})\left({\nu \over \nu_0}\right)^{-\beta(\hat{n})}\,,
\end{equation}
where $A_{\nu_0}$ is the observed foreground template at certain reference frequency $\nu_0$
and $\beta(\hat{n})$ denotes spectral index of the foreground component. A foreground for
which the approximation of a constant value of $\beta$ over the entire sky holds is said
to follow rigid frequency scaling, though, in a most general case $\beta$ is a function of
position of the sky, $\hat{n}$. If for a foreground, the assumption of constant $\beta$ is not
reasonable, one can model the variation in terms of two components each of which follows
rigid scaling (Bouchet and Gispert 1999). So, in that case,
\begin{equation}
 F(\hat{n}) = A_{\nu_0}\left({\nu \over \nu_0}\right)^{-\overline{\beta}} +
	      B_{\nu_0}(\hat{n})\left({\nu \over \nu_0}\right)^{-\overline{\beta}}\ln(\nu / \nu_0)\,,
\end{equation}
where $\overline{\beta}$ denotes average spectral index of the foreground, $F(\hat{n})$, over
the entire sky and $A,B$ are two templates of the foreground. A very strong variation may require
more than two templates for modelling.

Here, assuming a rigid scaling behaviour for the foreground component, we model the spectral coefficients
of the foreground in a frequency band $b$ as,
\begin{equation}
a_{lm}^{(f)b} = A_{lm}\left({1 \over f_b}\right)^{\alpha}\,.
\end{equation}
Now, Eq. \ref{alm_ea_term2-b} becomes,
\begin{equation}
{\langle a_{lm'}^{(s)}\hat{w}_l^{b'} a_{lm''}^{(f)b'*} \rangle}  = 
-A_{lm'}A_{lm''}^*{C_l^{(s)} \over (2l+1)\delta}\left({1 \over f_1^\alpha} - {1 \over f_2^\alpha} \right)^2\,.
\end{equation}

The third term after second equality in Eq. \ref{alm_ens_avg} is same as the second term. 
The fourth term is,
\begin{eqnarray}
 {\langle \hat{w}^{b}_l a_{lm'}^{(f)b} {a_{lm''}^{(f)b'}}^* \hat{w}^{b'}_l \rangle} &=&
{\left\langle \left(\hat{w}_l^1 a_{lm'}^{(f)1} + \hat{w}_l^2 a_{lm'}^{(f)2}\right)\left(\hat{w}_l^1 a_{lm''}^{(f)1} +
\hat{w}_l^2 a_{lm''}^{(f)2}\right)^* \right\rangle} \nonumber \\
&=& {\langle(\hat{w}_l^1)^2\rangle} a_{lm'}^{(f)1} a_{lm''}^{(f)1*} +
    2{\langle\hat{w}_l^1\hat{w}_l^2\rangle} a_{lm'}^{(f)1} a_{lm''}^{(f)2*} \nonumber \\
 && \,+\,{\langle(\hat{w}_l^1)^2\rangle} a_{lm'}^{(f)2} a_{lm''}^{(f)2*}\,.
\end{eqnarray}
With rigid scaling approximation for the foreground and after a lengthy algebra, we finally get,
\begin{equation}
 {\langle \hat{w}^{b}_l a_{lm'}^{(f)b} {a_{lm''}^{(f)b'}}^* \hat{w}^{b'}_l \rangle} =
 A_{lm'}A_{lm''}^*{C_l^{(s)} \over (2l+1)\delta}\left({1 \over f_1^\alpha} - {1 \over f_2^\alpha} \right)^2\,.
\end{equation}
Here we have used,
\[
\delta = C^{(f)1}_l + C^{(f)2}_l - 2C^{(f)1(f)2}_l = {1\over 2l+1}
\sum_{m=-l}^{m=+l}|A_{lm}|^2  \left({1 \over f_1^\alpha} - {1 \over f_2^\alpha} \right)^2\,.
\]
Finally we get from Eq. \ref{alm_ens_avg},
\begin{equation}
 {\langle a_{lm'}^{Clean} a_{lm''}^{Clean*} \rangle} = C_l^{(s)} \delta_{m'm''} -C_l^{(s)} A_{lm'}A_{lm''}^* \left(\sum_{m=-l}^{m=+l}|A_{lm}|^2\right)^{-1}
 \,.
\label{eq:lowLbias}
\end{equation}
Hence we find that the low-$l$ negative bias term leads to an anisotropic power 
tensor. As discussed in (Saha  et al. 2008; Samal et al. 2010) this bias arises 
even in the case of perfect cleaning, where we have enough frequency channels
to remove all the foregrounds.  
Besides this term we also expect a positive contribution due to 
residual 
foreground contamination. This arises due to foregrounds which may not
be eliminated by our cleaning procedure. By explicit numerical computation
we find that the PEV of the power tensor 
corresponding to Eq. \ref{eq:lowLbias} is well aligned with 
the galactic poles. The two orthogonal vectors of the power tensor lie 
in the galactic plane. This is of course expected since the eigenvectors
are determined entirely by the galactic foregrounds, which are present
predominantly along the galactic plane.

The main result of this section is that power tensor corresponding 
to the cleaned map acquires an anisotropic contribution due to low-$l$
negative bias. This contribution arises even in the case of perfect
cleaning when we have sufficient frequency channels to remove all the 
foregrounds present in the data. Although we have made several simplifying
assumptions, this basic result is expected to hold in general. 
Our analytic results in this section, however, do not provide any guidance
on how the PEV is affected due to foreground cleaning. 
In the next section we estimate the total
bias in the PEV by direct simulations.

\section{Results}
\subsection{Observational data}
\begin{figure}
 \centering
  \includegraphics[width=0.98\textwidth]{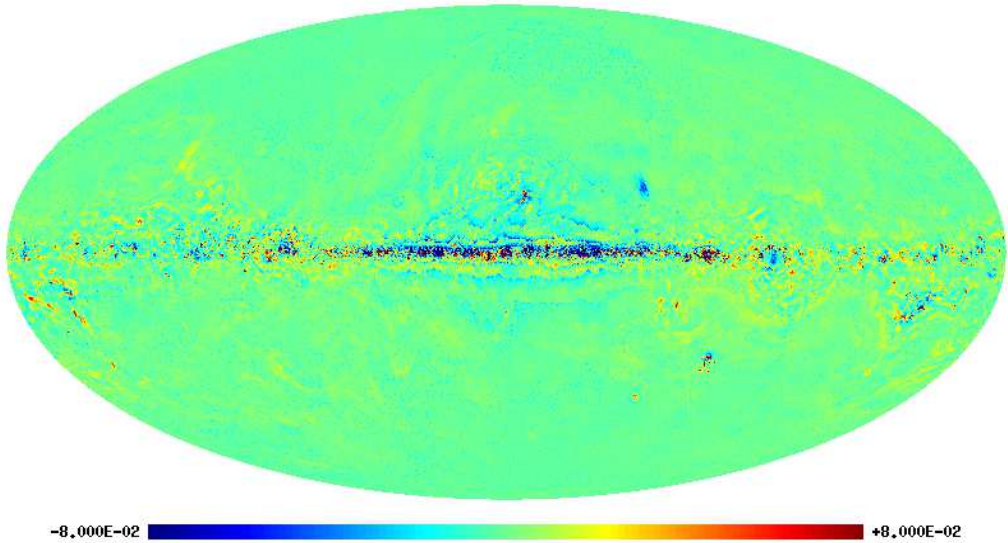}
  \caption{The bias map for seven year data
generated using the IPSE procedure.}
  \label{bias_map}
\end{figure}

\begin{figure}
 \centering
  \includegraphics[angle=-90,width=0.98\textwidth]{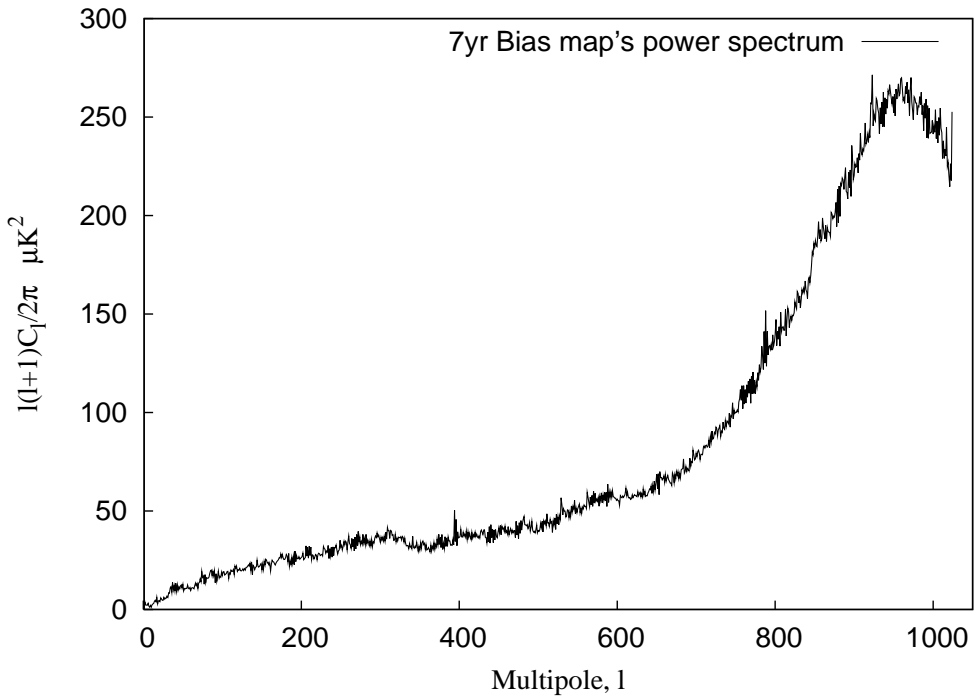}
  \caption{The power spectrum of the bias map shown in Fig. [\ref{bias_map}].}
  \label{bias_cl}
\end{figure}

We, first, present the results from actual cosmological data and then discuss
the simulation results. The PEVs, $\hat n_2$ and $\hat n_3$, 
for the quadrupole and octopole, respectively, from different clean maps extracted
from the WMAP data are given in Table [\ref{tab_PEV}]. The results presented here
are for the WMAP ILC maps and the IPSE cleaned maps, with and without bias
correction, for both five and seven year data sets. 
The bias map,
generated from Monte Carlo simulations of 500 random realizations, is
shown in Fig. [\ref{bias_map}]. The power spectrum corresponding
to this bias map is shown in Fig. [\ref{bias_cl}]. 
As expected, we see that the bias correction
is present, predominantly, in the galactic plane and is
almost zero away from it. We point out that this bias is
completely different from the negative bias discussed in the previous
section.

The ILC map is obtained by linearly combining the temperature maps,
observed in different frequency bands, in real space, with appropriate weights
to remove the foregrounds. In contrast, the IPSE map is obtained by making
linear combinations in harmonic space. We point out that, the bias removal leads
to a significantly improved cleaning in the galactic plane and has already been
applied in the publicly available ILC map. The angle, $\delta\Theta$, between
the two axes, $\hat n_2$ and $\hat n_3$, for different foreground
cleaned maps, is listed in
Table \ref{tab_angle}. Here, we also list the analytic estimate of probability
for the alignment found in WMAP CMBR data to be a random occurrence.

We find that, the alignment between quadrupole and octopole is striking in
both the ILC five and seven year maps. We also note that, in WMAP 3 year ILC map,
the alignment of quadrupole and octopole is found to be $5.51^o$, which has a
random occurrence probability of 0.0051. So, with more data release, the
alignment appears to be getting better. The significance for the observed
alignment in seven year data is better than $4\sigma$. The IPSE cleaned, bias
corrected 7 year map gives significance close to $3\sigma$. In comparison,
the IPSE map without bias correction gives a much poorer alignment with
$\delta\Theta = 11.28^o$(a 98.1\% CL). The main difference between these
two maps is that a significant foreground contamination in the galactic
plane has been removed during the process of bias removal. This suggests
that the effect of foregrounds is to distort the signal of alignment rather
than to cause alignment.

We also notice from Table[\ref{tab_PEV}] that, as we remove the bias in the 
IPSE map, the octopole PEV changes only by a negligible amount, whereas the 
quadrupole undergoes a significantly large change. Indeed, after bias removal,
the quadrupole PEV is relatively close to that corresponding to the ILC map.
The IPSE octopole PEV agrees well with that obtained from ILC map, irrespective
of whether the bias is removed or not. So, it seems that quadrupole is more
susceptible to distortion due to the presence of residual foregrounds.

\begin{table}\footnotesize
	\begin{center}
		\begin{tabular}{c c c}
		\hline
			& {$\hat n_2$} & {$\hat n_3$} \\
		\hline \hline
		WMAP ILC map (5 yr) & $(0.2458,0.4135,-0.8767)$ & $(0.2505,0.3823,-0.8895)$ \\
		WMAP ILC map (7 yr) & $(0.2484,0.3921,-0.8857)$ & $(0.2434,0.3842,-0.8906)$ \\
		IPSE cleaned map (5 yr) &  $(0.1481,0.1923,-0.9701)$ & $(0.2040,0.3863,-0.8996)$ \\
		Bias corrected IPSE map (5 yr) & $(0.2936,0.4091,-0.8640)$ & $(0.2086,0.3855,-0.8988)$ \\
		IPSE cleaned map (7yr) & $(0.1355,0.1957,-0.9713)$ & $(0.1782,0.3773,-0.9088)$ \\
		Bias corrected IPSE map (7 yr) & $(0.2301,0.3240,-0.9176)$ & $(0.1773,0.3754,-0.9097)$ \\
		\hline
		\end{tabular}
	\end{center}
\caption{The principal eigenvectors (PEVs) $\hat n_2$ and $\hat n_3$
for quadrupole and octopole, respectively, for different cleaned maps}
\label{tab_PEV}
\end{table}

\begin{table}\footnotesize
	\begin{center}
		\begin{tabular}{c c c}
		\hline
			& {$\delta \Theta$} & {$1-\cos(\delta\Theta)$} \\
		\hline \hline
		WMAP ILC map (5 yr) & $1.95^o$ & 0.00058 \\
		WMAP ILC map (7 yr) & $0.6^o$ & $5.5\times 10^{-5}$ \\
		IPSE cleaned map (5 yr) &  $12.27^o$ & 0.023 \\
		Bias corrected IPSE map (5 yr) & $5.44^o$ & 0.0045 \\
		IPSE cleaned map (7 yr) &  $11.28^o$ & 0.019 \\
		Bias corrected IPSE map (7 yr) & $4.25^o$ & 0.0027 \\
		\hline
		\end{tabular}
	\end{center}
\caption{Alignment between quadrupole and octopole for different cleaned maps.
Also listed are the analytic estimates of the
probability for a random occurrence of the observed alignments.}
	\label{tab_angle}
\end{table}

\subsection{Simulation Results}

We next use simulated CMB maps to determine the effect of foreground
cleaning on the properties of the extracted CMB signal. 
The simulated maps containing
CMB, foregrounds and detector noise are cleaned using the IPSE procedure. In each
of the simulated maps, the bias due to foreground power in the galactic plane
is removed in exactly the same manner as in the real data. Even after removing
this bias, we expect that the PEV extracted from the cleaned map may not be exactly
the same as that corresponding to the pure CMB map. 

\begin{figure}
\centering 
\includegraphics[angle=-90,width=0.98\textwidth]{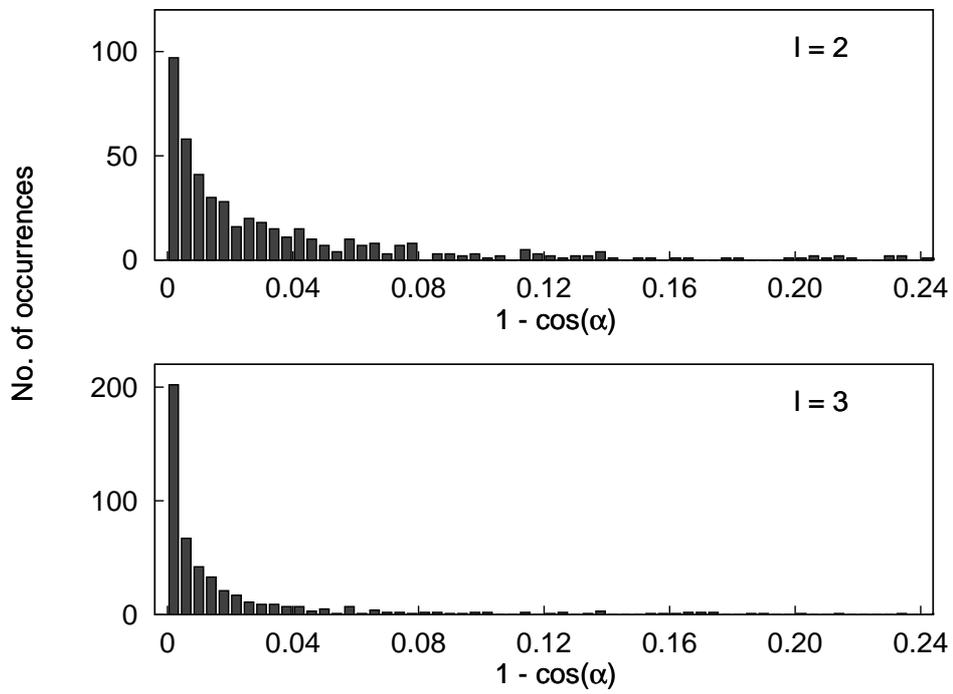} 
\caption{Distribution of $(1 - \cos\alpha)$, where $\alpha$ is the angle between the PEV
of simulated pure CMB maps and bias corrected
 cleaned maps for $l = 2$  and $l = 3$.} 
\label{hist_cos_alpha}
\end{figure}

Let $\alpha$ be the angle between
the pure map PEV ($ \hat{n}_p $) and cleaned map PEV ($ \hat{n}_c $) for a particular
multipole. Here, we are interested
only in the quadrupole and the octopole. We call this change in position of pure map
PEVs as ``rotation" of PEVs in the presence of foregrounds. We want to find out whether
there exists any preferred alignment of the pure map PEVs along a particular direction
in the presence of foregrounds. 
In Fig. [\ref{hist_cos_alpha}], we show the distribution
of ``$1-\cos(\alpha)$", which is a convenient measure of the rotation of
the PEVs. One can readily see from the graph that
most of the PEVs do not undergo any significant shift. In fact, the
number of PEVs which got rotated beyond a cutoff of $ 1 - \cos(\alpha) = 0.15 $ are 52 
for \emph{l}=2, and 30 for \emph{l}=3, out of the 500 bias corrected clean maps. Hence,
the error introduced due to foregrounds on the extraction of the PEVs is generally
small.

\begin{figure} 
\centering 
\includegraphics[angle=-90,width=0.9\textwidth]{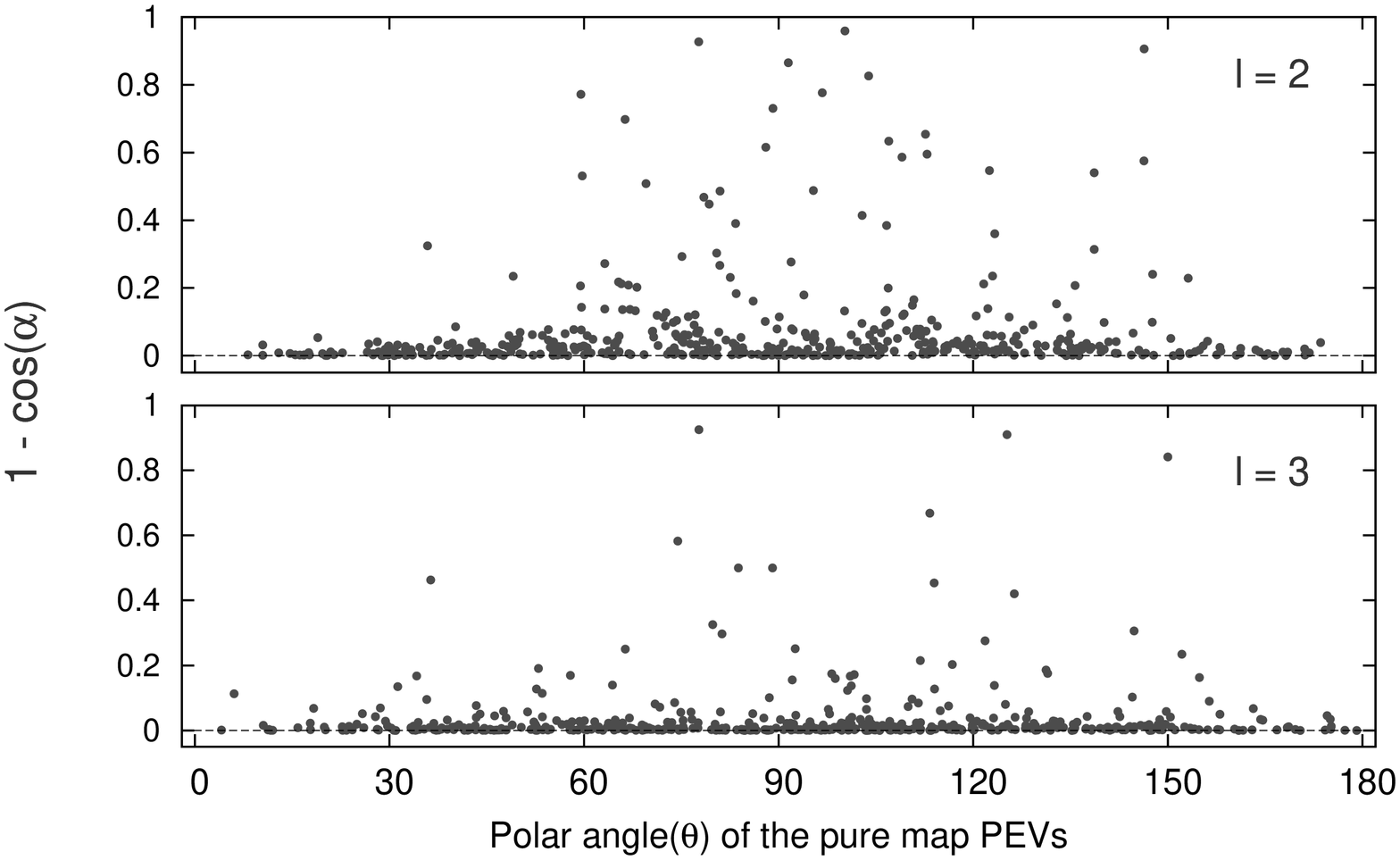} 
\caption{Plot of $(1-\cos\alpha)$, as a function of polar angle, $\theta$,
of pure map PEVs. Here, $\alpha$ is the angle between the PEVs of pure CMB maps
and bias corrected cleaned maps.} 
\label{cos_alpha_theta}
\end{figure}

\begin{figure} 
\centering
\includegraphics[angle=-90,width=0.9\textwidth]{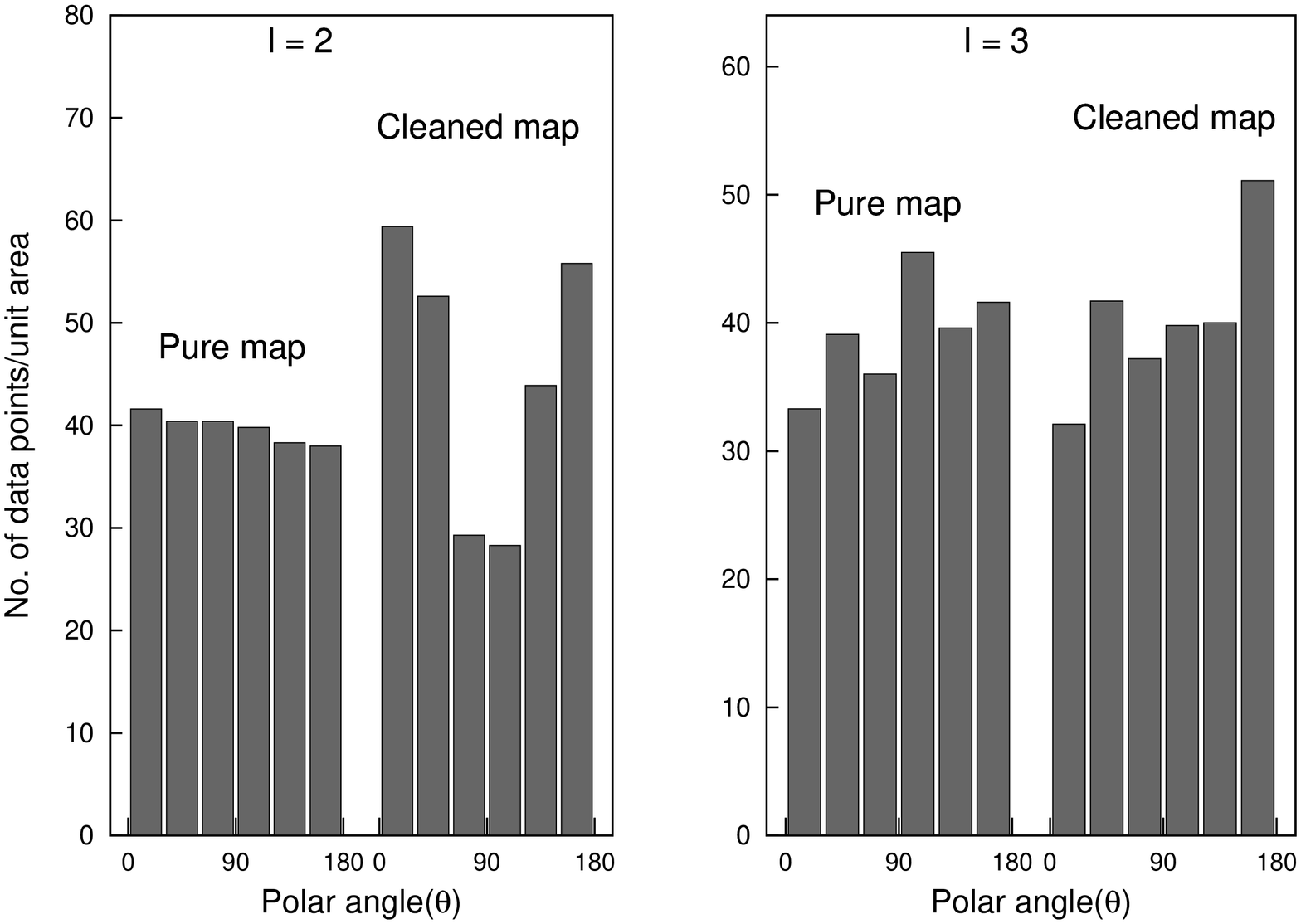} 
\caption{Distribution of the polar angle of PEVs for both the pure map and cleaned map, after bias correction,
for $l\,=\,2,\,3$.}
\label{hist_sky_dist_l2_l3}
\end{figure}

Fig. [\ref{cos_alpha_theta}] shows the plot of ``$1-\cos(\alpha)$", for all the 500
simulated maps, as a function of the polar angle, $\theta$, of the pure map PEV,
both for $l=2$ and $l=3$. Notice the clustering of data points near the zero of
y-axis in this plot, which indicates only a very nominal shift in most of the PEVs.
It also shows that the PEVs which were lying, initially, in the galactic plane
generally undergo larger rotation whereas the effect is small on PEVs 
at higher latitudes. Note that the galactic foregrounds extend roughly upto
$ 30^o $ on either side of the galactic equator and are distributed 
asymmetrically. Next, in Fig. [\ref{hist_sky_dist_l2_l3}]
we show the distribution of PEVs in different sky regions for $l = 2$ and $l = 3$. The
histograms are shown for both pure and cleaned maps in an interval of $30^o$ of
polar angle across the sky. The distribution of pure map PEVs is relatively flat,
as expected. However, the foreground cleaned maps show a dip near
the galactic plane. This illustrates that in the cleaned maps the PEVs have
a tendency to rotate away from the galactic plane. This effect is seen to be more
pronounced in the case of quadrupole in comparison to octopole. The effect is easily
understood since the region of the galactic plane is most contaminated by foregrounds.
Hence the PEVs in the galactic plane are naturally expected to undergo larger rotation.

\begin{figure} 
\centering 
\includegraphics[angle=-90,width=0.9\textwidth]{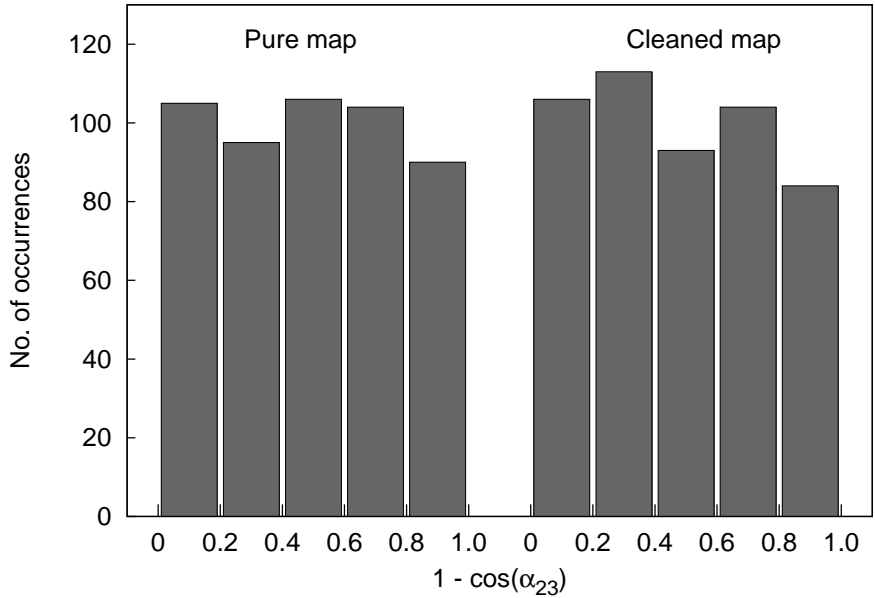} 
\caption{Distribution of $(1-\cos(\alpha_{23}))$ where $\alpha_{23}$ is
the angle between the quadrupole and octopole both for the pure map and 
the cleaned map, after bias correction.} 
\label{hist_align_l2_l3}
\end{figure}

\begin{figure}
\centering
\includegraphics[angle=-90,width=0.98\textwidth]{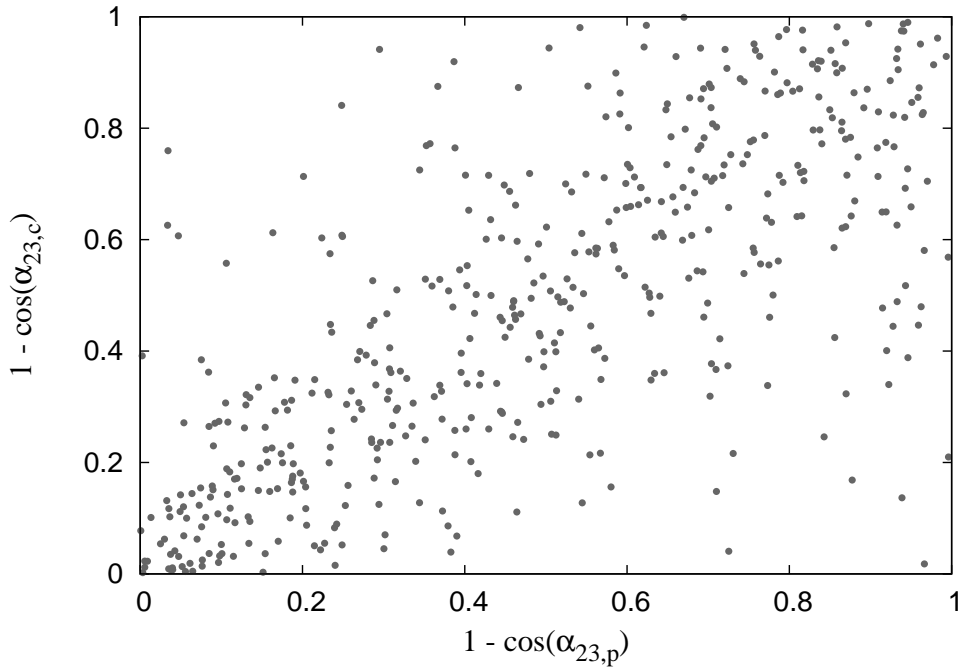}
\caption{The alignment between quadrupole and octopole of pure map and bias
corrected cleaned
maps are plotted here, to find any correlations between them. On y-axis, is
$1 - \cos(\alpha_{23,c})$, where $\cos \alpha_{23,c} = \hat{n}_{l2} \cdot \hat{n}_{l3}$
and $\hat{n}_{l2}$ and $\hat{n}_{l3}$ are the PEVs corresponding to $l = 2$ and $l = 3$,
respectively, for the simulated foreground cleaned maps. On x-axis we have the same
measure corresponding to the pure CMB maps.}
\label{align_l2_l3}
\end{figure}

Next, we look at the alignment of quadrupole and octople in both pure and cleaned
maps. Fig. [\ref{hist_align_l2_l3}] shows the distribution of ``$1-\cos(\alpha_{23})$",
where $\alpha_{23}$ is the angle between the quadrupole and the octopole. 
 The data is binned with
a bin size of $0.2$. We can easily infer from the figure that these alignments
are completely random. Hence, we find no evidence for foregrounds induced alignment
in the cleaned maps. In Fig. [\ref{align_l2_l3}] we show a plot of
$(1-\cos(\alpha_{23,p}))$ vs $(1-\cos(\alpha_{23,c}))$ for all the 500 simulations.
Here, the subscripts `$p\,$' and `$c\,$' stand for ``pure" and ``clean" simulated CMB maps,
respectively. If there is any correlation it will show up in such a plot. But, this plot
illustrates that there is no sign of preferred alignment.

We next compute the P-value or the probability that the alignment seen 
in data arises due 
to a random fluctuation directly from simulations. We determine the
number of random realizations of CMB data which show better alignment in
comparison to what is observed. 
The probability
is obtained by dividing this number by the total number of random samples. 
Here we use both the randomly generated
pure CMB maps as well as simulated foreground cleaned maps. 
For the bias 
corrected IPSE five year map we find that the probability is 0.004
if we use the simulated pure CMB maps. The probability
remains unchanged if it is computed by using the simulated
cleaned maps.
The corresponding numbers for the seven year maps are 0.004 and 0.002
for the simulated pure CMB maps and the 
cleaned maps respectively. The result remains unchanged if we apply the residual
foreground bias correction to the simulated cleaned maps.  
These results agree well
with those given in Table 2.  

Hence we find no evidence for any systematic
effect which may cause alignment of PEVs, despite the
presence of a systematic effect on power, $ C_l $. We do find a systematic
shift of the PEVs away from the galactic plane. However the final PEVs 
point in random directions. Hence rather than causing alignment, this
effect has a tendency to distort any signal of alignment that might be
present in the original sample. 

\begin{figure} 
\centering 
\includegraphics[angle=-90,width=0.9\textwidth]{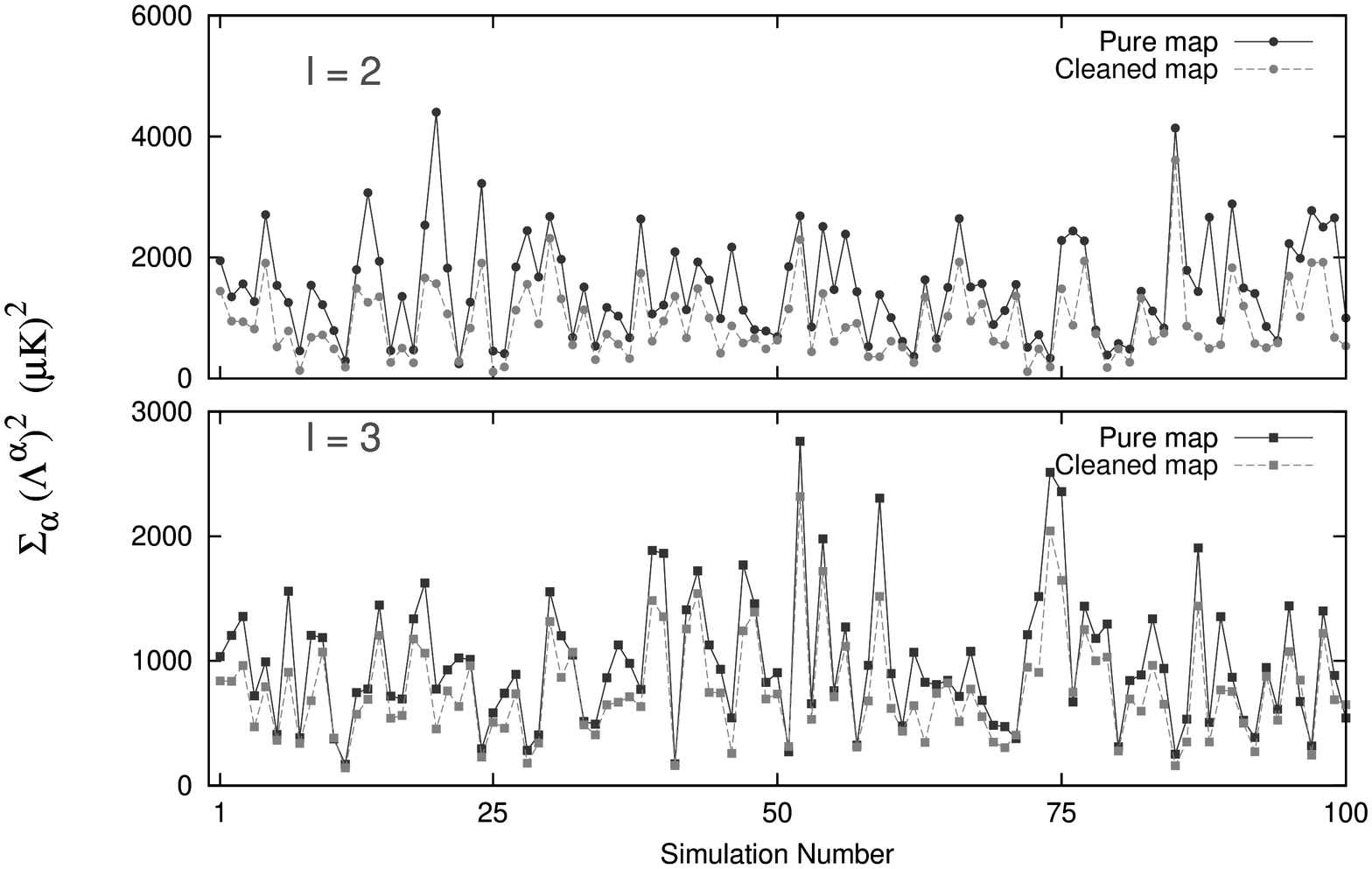} 
\caption{The sum of eigenvalues, or the total power, for $l=2,3$ for the
simulated pure CMB and the cleaned CMB map, after bias correction. 
We see a systematic
negative shift in the power for each of the simulated maps.
} 
\label{sum_eigen_l2}
\end{figure}

As we have mentioned earlier, the negative bias does cause a systematic reduction of power for low $l$ 
multipoles. This is seen clearly in Fig. 
[\ref{sum_eigen_l2}] where we plot the sum of the eigenvalues of the 
power tensor for $l=2,3$
both for simulated pure and cleaned, bias corrected, maps. The sum
of eigenvalues equals the total power, $C_l$, for each multipole. We find
a systematic negative shift in the power for each random realization. 
After taking the average over all the simulations, 
we find that quadrupole power of the cleaned map is lower by 39 \% 
in comparison to that of the pure CMB map. 
The corresponding octopole power gets reduced by 22 \%. 

We next assume that the alignment is caused by some effect other than
foreground contamination. The foregrounds lead to a random shift
in the PEVs causing a certain level of misalignment. Hence the angle
between $l=2$ and $l=3$ PEVs must not be very small in comparison 
to the shift caused by foregrounds. 
The shift of the PEVs is quantified by the distribution plot shown in 
Fig. [\ref{hist_cos_alpha}]. This should be associated with the randomness
that the cleaning process introduces in the extraction of the PEV. 
We find that probability of chance alignment for the IPSE cleaned
five year 
and seven year map, after bias correction, is $0.0045$ and $0.0027$
respectively. We find that about 16\% of the
simulations have $(1-\cos\alpha)<0.0027$. At high galactic latitudes,
corresponding to the Virgo alignment axis, this 
number is higher. It is reassuring that the number of random
samples which show a shift in the PEV less than the observed alignment in the
IPSE cleaned map is not too small.

\section{Conclusions}
We have studied the effect of foreground cleaning on the alignment of low
$l$ multipoles, $l=2,3$. It has been speculated in the past that the
alignment may be caused by foreground contamination. The presence of
residual foreground power in the cleaned CMB maps could lead to violation
of statistical isotropy and hence alignment of low $l$ multipoles. 
Furthermore, even in the case of perfect cleaning, the extracted power
contains a negative bias, which is quite significant at low $l$. This has
been shown explicitly in the case of the IPSE procedure. We anticipate
the presence of such a bias also in the ILC map but this requires further 
study. Using the IPSE procedure, we analytically compute the power
tensor using some simplifying assumptions. We find that its ensemble 
average is not proportional to an identity matrix due to the presence
of the low $l$ bias. Hence this implies a violation of statistical isotropy.  
We next perform detailed numerical simulations within the framework of the
IPSE method to determine 
the effect of foregrounds on the observed alignment.

Our IPSE simulation results, obtained using 500 randomly generated samples,
show that the PEVs extracted from foreground cleaned maps are generally
in good agreement with those obtained from pure CMB maps. In most cases
the shift in PEVs is observed to be small. The largest change is seen
in the case of quadrupole. The shift is dominant for PEVs which initially
lie close to the galactic plane. In general, we find 
that the PEVs tend to get pushed away from the galactic plane
towards the galactic poles. Similar trend is also seen in the case
of octopole but with much reduced significance. We do not, however, see
any evidence for alignment among the quadrupole and octopole PEVs 
extracted from the foreground cleaned maps. The significance or P-values
extracted by using simulated cleaned
maps are found to be in good agreement with those obtained by using
simulated pure CMB maps. These also show good agreement with the
results obtained analytically and listed in Table 2.     

Hence despite the systematic shift of PEVs away from the galactic plane, 
as observed dominantly in the case of quadrupole, we find that foregrounds
do not cause alignment of quadrupole and octopole. This is because 
the change in PEVs due to foreground cleaning is generally small and
 mostly in random directions. One might be tempted to propose a scenario
where the shift towards galactic poles is a much more significant effect
which may be caused by the presence of much stronger level of foreground
contamination. In this case it is conceivable that the PEVs extracted from 
the foreground cleaned maps would point dominantly towards the galactic
poles and hence show alignment. However this would require a very large foreground contamination
in comparison to currently accepted model. The level of contamination has to 
be sufficiently large so that both the quadrupole and octopole PEVs 
tend to align with the galactic poles. Although the quadrupole shows
a somewhat significant shift of PEVs away from galactic plane, this trend
is barely noticeable in the case of octopole. We emphasize that
even in the case of quadrupole the shift of PEV is very small in most
cases. Furthermore such a proposal will imply
that the direction of alignment points towards the 
galactic poles which is not consistent
with the observed axis. Hence we conclude that foreground 
contamination is very unlikely to cause the observed alignment. 


Due to the random nature of the shift caused by foreground cleaning, it is
more likely to cause misalignment of multipoles. We see this clearly
by comparing the results using IPSE before and after removing the bias
due to foreground contamination in the galactic plane. The alignment is
found to be much stronger after removing the residual foreground bias 
which arises due to remnant foregrounds, present in our cleaned maps.
It is extracted by simulations using the PSM for foregrounds
and then removed from the cleaned map. Furthermore we see that alignment
becomes much more significant as more data is accumulated. The alignment
seen in seven year data, for example, is much better than seen in five year
data which in turn is better than that seen in three year data. 
This is clearly caused by reduced uncertainties  
as more data accumulates. Hence we conclude that both the foregrounds
and detector noise are more likely to distort, rather than cause, 
the signal of alignment.

\bigskip
\noindent
{\bf \large Acknowledgements:}
P. K. Samal thanks NISER,
Bhubaneswar for providing their computer and library facility.

\bigskip

{\bf\large  References}

\begin{itemize}

\item[] Abramo ~L.~R., Sodre ~Jr.,~L., Wuensche ~C.~A., 2006, Phys. Rev., D 74, 083515

\item[] Ackerman L., Carroll S. M., Wise M. B., 2009, Phys. Rev., D 75, 
069901

\item[] Armendariz-Picon ~C., 2004, JCAP, 7, 7

\item[] Battye ~R.~A.,  Moss ~A., 2006, Phys. Rev., D 74, 041301

\item[] Bennett C. L. et al., 2003, ApJS, 148, 1

\item[] Berera ~A., Buniy ~R.~V.,  Kephart ~T.~W., 2004, JCAP, 10, 16

\item[] Bernui ~A., Villela ~T., Wuensche ~C.~A., Leonardi ~R.,  
Ferreira ~I., 2006, A\&A, 454, 409

\item[] Bernui ~A., Mota ~B.,  Reboucas ~M.~J.,  Tavakol ~R., 2007, A\&A, 464, 479

\item[] Bielewicz ~P., Gorski ~K.~M., Banday ~A.~J., 2004, MNRAS, 355, 1283

\item[] Bielewicz ~P., Eriksen ~H.~K., Banday ~A.~J., Gorski ~K.~M., Lilje 
~P.~B., 2005, ApJ, 635, 750

\item[] Birch ~P., 1982, Nature, 298, 451

\item[] Bouchet ~F.~R., Gispert ~R., 1999, New Astronomy, 4, 443

\item[] Buniy ~R.~V., Berera ~A., Kephart ~T.~W., 2006, Phys. Rev., D 73, 063529

\item[] Carroll S. M., Tseng C.-Y., Wise M. B., 2010, Phys. Rev., D 81, 
083501

\item[] Copi ~C.~J., Huterer ~D., Schwarz ~D.~J., Starkman ~G.~D., 2006, MNRAS, 367, 79

\item[] Copi ~C.~J., Huterer ~D., Schwarz ~D.~J., Starkman ~G.~D., 2007, Phys. Rev., D 75, 023507

\item[] Cruz ~M., Martinez-Gonzalez ~E., Vielva ~P., Cayon ~L., 2005, MNRAS, 356, 29

\item[] de Oliveira-Costa ~A., Tegmark ~M., Zaldarriaga ~M., Hamilton ~A., 2004, Phys. Rev., D 69, 063516

\item[] de~Oliveira-Costa ~A.,  Tegmark ~M., 2006, Phys. Rev., D 74, 023005

\item[] Delabrouille J., Cardoso J. F., 2009, Data Analysis in Cosmology, Lecture notes in physics 665, eds Vicent J. Martinez et al. (Springer), 
pages 159-205

\item[] Erickcek A. L., Hirata C. M., Kamionkowski M., 2009, Phys. Rev., D 
80, 083507

\item[] Eriksen ~H.~K., Hansen ~F.~K., Banday ~A.~J., Gorski ~K.~M., Lilje 
~P.~B., 2004, ApJ, 605, 14

\item[] Eriksen ~H.~K., Banday ~A.~J., Gorski ~K.~M., Hansen ~F.~K., 
Lilje ~P.~B., 2007, ApJ, 660, L81

\item[] Freeman ~P.~E., Genovese ~C.~R., Miller ~C.~J., Nichol ~R.~C., 
Wasserman ~L., 2006, ApJ, 638, 1

\item[] Gold  B. et.al., 2010, arXiv:1001.4555

\item[] Gordon ~C., Hu ~W., Huterer ~D., Crawford ~T., 2005, Phys. Rev., D 72, 103002

\item[] Hajian ~A., Souradeep ~T., Cornish ~N., 2005, ApJ, 618, L63

\item[] Hajian ~A., Souradeep ~T., 2006, Phys. Rev., D 74, 123521

\item[] Hansen ~F.~K., Banday ~A.~J.,  Gorski ~K.~M., 2004, MNRAS, 354, 641

\item[] Helling ~R.~C., Schupp ~P.,  Tesileanu ~T., 2006, Phys. Rev., D 74, 063004

\item[] Hill R. S. {\it et al}, 2008, arXiv:0803.0570

\item[] Hinshaw ~G. et al., 2007, ApJS, 170, 288

\item[] Hoftuft ~J., Eriksen ~H.~K., Banday ~A.~J., Gorski ~K.~M., Hansen ~F.~K.,  Lilje ~P.~B., 2009, ApJ, 699, 985

\item[] Hutsem\'{e}kers ~D., 1998, A\&A, 332, 410

\item[] Hutsem\'{e}kers ~D.,  Lamy ~H., 2001, A\&A, 367, 381

\item[] Inoue ~K.~T.,  Silk ~J., 2006, ApJ, 648, 23

\item[] Itoh Y., Yahata K., Takada M., 2009, arXiv:0912.1460   

\item[] Jain ~P.,  Ralston ~J.~P., 1999, Mod. Phy. Lett. A, 14, 417

\item[] Jain ~P., Narain ~G.,  Sarala ~S., 2004, MNRAS, 347, 394

\item[] Kahniashvili T., Lavrelashvili G., Ratra B., 2008,
Phys. Rev., D 78, 063012 
 
\item[] Kashlinsky ~A., Atrio-Barandela ~F., Kocevski ~D.,  Ebeling ~H., 2008, ApJ, 686, L49

\item[] Kashlinsky ~A., Atrio-Barandela ~F., Kocevski ~D.,  Ebeling ~H., 2009, ApJ, 691, 1479

\item[] Katz ~G.,  Weeks ~J., 2004, Phys. Rev., D 70, 063527

\item[] Koivisto ~T.,  Mota ~D.~F., 2008, ApJ, 679, 1

\item[] Land ~K.,  Magueijo ~J., 2006, MNRAS, 367, 1714

\item[] Land ~K.,  Magueijo ~J., 2007, MNRAS, 378, 153

\item[] Magueijo ~J., Sorkin ~R.~D., 2007, MNRAS, 377, L39

\item[] Moffat ~J.~W., 2005, JCAP, 10, 12

\item[] Naselsky ~P.~D., Verkhodanov ~O.~V., Nielsen ~M.~T.~B., 2008, Astrophys. Bull., 63, 216

\item[] Prunet ~S., Uzan ~J., Bernardeau ~F., Brunier ~T., 2005, Phys. Rev., D 71, 083508

\item[] Raki\'{c} ~A., Rasanen ~S., Schwarz ~D.~J., 2006, MNRAS, 369, L27

\item[] Raki\'{c} A., Schwarz D. J., 2007, Phys. Rev., D 75, 103002

\item[] Ralston ~J.~P., Jain ~P., 2004, Int. J. of Mod. Phys., D 13, 1857

\item[] Rodrigues ~D.~C., 2008, Phys. Rev., D 77, 023534

\item[] Saha ~R., Prunet ~S., Jain ~P., Souradeep ~T., 2008, Phys. Rev., D 78, 023003

\item[] Saha ~R., Jain ~P., Souradeep ~T., 2006, ApJL 645, L89 

\item[]  Samal P., Saha R.,  Delabrouille J., Prunet S., 
Souradeep T., 2010,  ApJ,  714, 840 

\item[] Samal ~P.~K., Saha ~R., Jain ~P., Ralston ~J.~P., 2008, MNRAS, 385, 1718

\item[] Samal ~P.~K., Saha ~R., Jain ~P., Ralston ~J.~P., 2009, MNRAS, 396, 511

\item[] Sarkar D., Huterer  D.,  Copi C. J.,  Starkman G. D., Schwarz D. J., 
2010, arXiv:1004.3784 

\item[] Schwarz ~D.~J., Starkman ~G.~D., Huterer ~D., Copi ~C.~J., 2004, PRL, 93, 221301

\item[] Slosar ~A., Seljak ~U., 2004, Phys. Rev., D 70, 083002

\item[] Tegmark M., Efstathiou, G., 1996, MNRAS 281, 1297 

\item[] Tegmark M., de Oliveira-Costa A,  Hamilton A. J., 2003,
Phys. Rev., D 68, 123523 

\item[] Wiaux ~Y., Vielva ~P., Martinez-Gonzalez ~E., Vandergheynst ~P., 2006, PRL, 96, 151303

\end{itemize}
\end{document}